\documentstyle[12pt]{article}

\begin{document}
\hfill\vbox{
\hbox{NUHEP-TH-95-10}
\hbox{hep-ph/9508406}
\hbox{August 1995} }\par
\thispagestyle{empty}

\begin{center}
{\Large \bf On the Convergence of Perturbative QCD \\
	 at High Temperature}

\vspace{0.15in}

Eric Braaten\footnotemark[1] and Agustin Nieto\footnotemark[1] \\

{\it Department of Physics and Astronomy, Northwestern University,
Evanston, IL 60208}

\end{center}

\begin{abstract}

The free energy for QCD at high temperature $T$ is calculated to order
$g^5$ using effective-field-theory methods to separate
the contributions from the momentum scales $T$ and $gT$.
The effects of the scale $T$ enter through the coefficients
in the effective lagrangian for the 3-dimensional effective theory obtained
by dimensional reduction.  The perturbation series for these coefficients
seem to be well-behaved if the running coupling constant is sufficiently
small:  $\alpha_s(2 \pi T) \ll 1$.  For the contribution to the
free energy from the scale $gT$, the perturbation series is well-behaved
only if $\alpha_s(2 \pi T)$ is an order of magnitude smaller.
The implication for applications of perturbative QCD to the quark-gluon
plasma are briefly discussed.
\end{abstract}

\footnotetext[1]{Address after August 1995: {\it Department of Physics,
Ohio State University, Columbus OH 43210}.}

\newpage

One of the most dramatic predictions of quantum chromodynamics (QCD)
is that when hadronic matter is raised to a sufficiently high temperature
or density, it will undergo a phase transition to a quark-gluon plasma.
One of the major thrusts of nuclear physics in the next decade will
be the  effort to study the quark-gluon plasma through
relativistic heavy-ion collisions.  For this effort to be successful,
it will be important to understand the
properties of the plasma as accurately as possible.
The two major theoretical tools for studying the quark-gluon plasma
are lattice gauge theory and perturbative QCD.  Lattice gauge theory
has the advantage that it is a nonperturbative method and applies equally
well to the hadronic phase.  It is an effective method
for calculating the static equilibrium properties of a plasma with 0
baryon density, but it can not be easily applied to problems involving
dynamical properties or to a plasma that is away from equilibrium or
has nonzero baryon density.  These problems can however be studied using
perturbative QCD, provided that the temperature $T$ of the plasma is
sufficiently high.  As the temperature decreases, the
running coupling constant $g$ of QCD increases, causing perturbation theory
to break down at some temperature above the
critical temperature $T_c$ for the phase transition.
One of the basic questions in the theory of the quark-gluon plasma
is how large must $T$ be in order for perturbative QCD to be applicable.
Is this method useful at temperatures that may be achievable in
heavy-ion collisions, which are at most several times $T_c$?

In order to answer this question, it is necessary to understand the
structure of the perturbation series to all orders and also to
carry out explicit higher-order calculations.
The first step has been carried out for the free energy \cite{solution}.
The structure of the perturbation series is nontrivial,
because a strict perturbation expansion
in $g^2$ has severe infrared divergences associated
with the exchange of static
gluons.  Physically, these divergences are screened by plasma effects.
The screening of electrostatic gluons can be taken into account by a
resummation of perturbation theory, but the screening of magnetostatic
gluons can only be treated using nonperturbative methods.
Once the structure of the perturbation expansion is understood,
it is still necessary to carry out explicit perturbative calculations
to determine quantitatively how high the temperature must be in order
for perturbative QCD to be accurate.
Only recently has the calculational technology of thermal field theory
progressed to the point where it is possible to carry out
perturbative calculations to a high enough
order that the running of the coupling
constant comes into play [2--8].
The first such calculation for the quark-gluon plasma
was the calculation of the free energy to order $g^4$ by Arnold and
Zhai in 1994 \cite{arnold}.  This calculation has recently been
extended to order $g^5$ by Kastening and Zhai \cite{zhai}
and by the authors \cite{eftQCD}.  In this Letter,
we summarize the calculations of Ref. \cite{eftQCD}
and discuss implications for the application of
perturbative QCD to the quark-gluon plasma.

The static equilibrium properties of a quark-gluon plasma
at temperature $T$ are governed by the free energy density
$F = - (T/V) \log {\cal Z}_{\rm QCD}$,
where $V$ is the volume of space.  The partition function
${\cal Z}_{\rm QCD}$
is given by a functional integral over quark and gluon fields
on a 4-dimensional Euclidean space-time, with the Euclidean time
$\tau$ taking its values on a circle with circumference $1/T$.
In the limit in which the quarks are massless, the free energy is a
function of $T$ and $\alpha_s = g^2/(4 \pi)$ only.

The structure of the weak-coupling expansion for the free energy
to all orders was deduced in Ref. \cite{solution}.
The free energy can be expressed as the sum of three contributions
coming from the momentum scales $T$, $gT$, and $g^2T$:
\begin{equation}
F \;=\;  \left[ f_E(\Lambda_E) \;+\; f_M(\Lambda_E,\Lambda_M)
	\;+\; f_G(\Lambda_M) \right] T \;,
\label{free}
\end{equation}
where $\Lambda_E$ is an arbitrary factorization scale that
separates the scales $T$ and $gT$, while
$\Lambda_M$ separates the scales $gT$ and $g^2T$.
The contributions from the three momentum scales can
be unraveled by constructing a sequence of two effective field theories.
The first effective theory, electrostatic QCD (EQCD),
is a 3-dimensional Euclidean field theory involving the
electrostatic gauge field $A_0^a({\bf x})$
and the magnetostatic gauge field $A_i^a({\bf x})$.
The lagrangian for EQCD is
\begin{equation}
{\cal L}_{\rm EQCD} \;=\;
{1 \over 4} G^a_{ij} G^a_{ij} \;+\; {1 \over 2} (D_i A_0)^a (D_i A_0)^a
\;+\; {1 \over 2} m_E^2 A_0^a A_0^a
\;+\; \delta {\cal L}_{\rm EQCD},
\label{LEQCD}
\end{equation}
where $D_i$ is the covariant derivative
for the adjoint representation with coupling constant $g_E$
and $G^a_{ij}$ is the magnetostatic field strength.
The term $\delta {\cal L}_{\rm EQCD}$
in (\ref{LEQCD}) includes all other local gauge-invariant operators
that can be constructed out of $A_0$ and $A_i$,
including nonrenormalizable interactions.
Static gauge-invariant correlation functions in thermal QCD
can be reproduced at long distances by tuning $g_E$, $m_E^2$,
and the parameters in $\delta {\cal L}_{\rm EQCD}$ as functions of
$g$, $T$, and the ultraviolet cutoff $\Lambda_E$ of EQCD.
In physical quantities,
the $\Lambda_E$-dependence of the parameters cancels the
$\Lambda_E$-dependence from loop integrals in EQCD.
Since the parameters of EQCD take into account effects from the scale $T$,
they can be calculated as perturbation series
in the running coupling constant $\alpha_s(\mu)$ with $\mu$ of order $T$.

Since the free energy is a static quantity, it can be calculated
using EQCD.  The free energy can be written
\begin{equation}
F \;=\; T \; \left( f_E
	\;-\; {\log {\cal Z}_{\rm EQCD} \over V} \right)\,,
\label{FEQCD}
\end{equation}
where ${\cal Z}_{\rm EQCD}$ is the partition function for EQCD
and $f_E$ is the coefficient of the unit operator, which was omitted from
the effective lagrangian (\ref{LEQCD}).  This coefficient
gives the contribution to the free energy (\ref{free}) from
the momentum scale $T$.  The logarithm of ${\cal Z}_{\rm EQCD}$
in (\ref{FEQCD}) includes the contributions $f_M$ and $f_G$
from the scales $gT$ and $g^2T$, respectively.
These contributions can be separated by constructing a second effective
field theory, magnetostatic QCD (MQCD), which involves only
the magnetostatic gauge field $A^a_i({\bf x})$.
The term $f_M$ in (\ref{free}) is the coefficient
of the unit operator in the lagrangian for MQCD.  It
can be computed using perturbative methods as an expansion
in powers of $g$ starting at order $g^3$.
The term $f_G$ in (\ref{free}) is proportional to the
logarithm of the partition function of MQCD.
It can only be calculated using nonperturbative methods.  Surprisingly,
it can be expanded in powers of $g$ beginning at order $g^6$,
with coefficients that can be calculated
using lattice simulations of MQCD \cite{solution}.  Since we only
calculate the free energy to order $g^5$, we do not consider the
term $f_G$ any further.

To calculate the free energy to order $g^5$, the only parameters of EQCD
that are required are  $g_E^2$ to leading order in $g^2$ and $m_E^2$ and
$f_E$ to order $g^4$.  The gauge coupling constant $g_E$ for EQCD
is determined at leading order simply by comparing the lagrangians
for EQCD and full QCD:  $g_E^2 = g^2 T$.
The other two parameters can be determined by computing
static quantities in both full QCD and EQCD,
and demanding that they match.
It is convenient to carry out these matching calculations
using a strict perturbation expansion in $g^2$.
This expansion is afflicted with infrared
divergences due to long-range forces mediated by static gluons,
and an infrared cutoff is therefore required.  Physically,
these divergences are screened by plasma effects, but screening
is not taken into account in the strict perturbation expansion.
Nevertheless, this expansion can be used as a device for determining
the parameters of EQCD, since they depend only on short distances
of order $1/T$.

The parameter $m_E$ can be determined by matching the
strict perturbation expansions for the electric screening mass $m_{\rm el}$
in full QCD and in EQCD.  Beyond leading order in $g$, $m_{\rm el}$
becomes sensitive to magnetostatic screening
and requires a nonperturbative definition \cite{yaffe}.
However, in the presence of an infrared cutoff,
the electric screening mass
can be defined in perturbation theory by the location of the pole in the
propagator for $A_0(\tau,{\bf x})$ at spacelike momentum
$(k_0=0,{\bf k})$.  Denoting the appropriate component of the gluon
self-energy tensor by $\Pi(k^2)$ where $k^2 = {\bf k}^2$, we must solve
the equation $k^2 \;+\; \Pi(k^2) = 0$ at $k^2=-m_{\rm el}^2$.
Since the solution $m_{\rm el}^2$ is of order $g^2$, we can expand
$\Pi(k^2)$ as a Taylor series around $k^2=0$.
To determine $m_{\rm el}^2$ to order $g^4$, we must calculate $\Pi(0)$
to two-loop accuracy and $\Pi'(0)$ to one-loop accuracy.
We use dimensional regularization with $3-2 \epsilon$ spacial dimensions
to cut off both infrared and ultraviolet
divergences.  The sums and integrals can be evaluated analytically
using methods developed by Arnold and Zhai \cite{arnold}.
The resulting expression for $m_{\rm el}^2$
is an expansion in integral powers of $\alpha_s$.
There is no $\alpha_s^{3/2}$ term, unlike in the expression for $m_{\rm el}^2$
that correctly incorporates the effects of electrostatic screening
\cite{rebhan}.  This $g^3$ term
arises because the $g^4$ correction includes a linear
infrared divergence that is cut off at the scale $gT$.
Since we use dimensional regularization as an infrared cutoff,
this power infrared divergence is set equal to 0.

In EQCD with an infrared cutoff, the electric
screening mass $m_{\rm el}$ can be defined in perturbation theory
by the location of the pole in the
propagator for the field $A_0({\bf x})$.  Denoting the self-energy
function for $A_0({\bf x})$ by $\Pi_E(k^2)$,
the screening mass $m_{\rm el}$ satisfies
$k^2 + m_E^2 + \Pi_E(k^2) = 0$ at $k^2 = -m_{\rm el}^2$.
In the strict perturbation expansion for EQCD, we treat
$m_E^2$ as a perturbation parameter of order $g^2$.
After Taylor-expanding $\Pi_E(k^2)$ around $k^2=0$, there
is no scale in the loop integrals, so they all vanish with
dimensional regularization.
The solution for the screening mass is therefore trivial:
$m_{\rm el}^2 =  m_E^2$.
Matching this result with the strict perturbation expansion from the full
theory and taking the limit $\epsilon \to 0$, we find
\begin{eqnarray}
m_E^2 \Bigg|_{\epsilon = 0} \;=\;
4 \pi \; \alpha_s(\mu) \; T^2
\Bigg\{ 1 + \textstyle{1 \over 6} n_F
\;+\; \Bigg[ 0.612 - 0.488 n_F - 0.0428 n_F^2
\nonumber \\
\;+\; {11 \over 2} \left(1 + {\textstyle{1 \over 6}} n_F \right)
	\left(1 - {\textstyle{2 \over 33}} n_F \right)
	\log {\mu \over 2 \pi T} \Bigg]
	{\alpha_s \over \pi}
\Bigg\} \;,
\label{mE}
\end{eqnarray}
where $n_F$ is the number of flavors of quarks and
$\mu$ is the renormalization scale for the QCD coupling constant.
The order--$\epsilon$ terms in $m_E^2$ are also required in
the calculation of the free energy.  These terms are given by
\begin{equation}
{\partial m_E^2 \over \partial \epsilon} \Bigg|_{\epsilon = 0}
\;=\; g^2 T^2 \Bigg\{
3.97 + 2 \log {\Lambda_E \over 4 \pi T}
\;+\; \left( 0.597 + {1 \over 3} \log {\Lambda_E \over 4 \pi T} \right) n_F
\Bigg\} ,
\label{dmEdeps}
\end{equation}
where the infrared cutoff $\Lambda_E$ is the scale introduced by dimensional
regularization.

The coefficient $f_E$ can be determined by matching the strict perturbation
expansions for the free energy in full QCD and in EQCD.
In full QCD, the free energy $F$ is calculated to order $g^4$
by evaluating the sum of vacuum diagrams through three-loop order,
using dimensional regularization
to cut off both infrared and ultraviolet divergences.
The resulting expression for $F$ is an expansion in
integral powers of $\alpha_s$.
There is no $\alpha_s^{3/2}$ term,
in contrast to the expression for the free energy that correctly
includes the effects of electrostatic screening \cite{shuryak,kapusta}.
This $g^3$ term arises
because the $g^4$ correction includes a linear infrared divergence
that is cut off at the scale $gT$.  In the strict perturbation expansion,
this term appears as a power infrared divergence that is set to zero
in dimensional regularization.

In EQCD, the free energy is given by (\ref{FEQCD}).
All the loop diagrams in the strict perturbation expansion
for $\log {\cal Z}_{\rm EQCD}$ vanish with dimensional regularization,
since there is no scale for the integrals.  The only contribution to
$\log {\cal Z}_{\rm EQCD}$ comes from the counterterm
$\delta f_E$ which cancels logarithmic ultraviolet
divergences proportional to the unit operator.
The resulting expression for the free energy is simply
$F = ( f_E + \delta f_E )T$.
The counterterm is determined by calculating the ultraviolet
divergent terms in $\log {\cal Z}_{\rm EQCD}$.
If we use dimensional regularization together with a
minimal subtraction renormalization scheme in EQCD,
then $\delta f_E$ is a polynomial in $g_E^2$, $m_E^2$,
and the other parameters in the EQCD lagrangian.
The leading term in $\delta f_E$ is proportional to $g_E^2 m_E^2$, and
its coefficient can be determined by a simple 2-loop calculation:
\begin{equation}
\delta f_E =
- {3 \over 8 \pi^2 \epsilon} g_E^2 m_E^2 .
\label{deltafE}
\end{equation}
When this counterterm is expressed in terms of the parameters $g$ and $T$
of the full theory, we must take into account the fact that $m^2_E$
in (\ref{deltafE}) multiplies a pole in $\epsilon$.
Thus, in addition to the expression for $m_E^2$
given in (\ref{mE}), we must also include the terms of order
$\epsilon$ which are given by (\ref{dmEdeps}).
Matching $F = (f_E + \delta f_E) T$ with the strict perturbation
expansion for $F$ in the full theory, we obtain
\begin{eqnarray}
f_E(\Lambda_E) \;=\;
- {8 \pi^2 \over 45} T^3
\Bigg\{ 1 + {\textstyle{21 \over 32}} n_F
\;-\; {15 \over 4} \left(1 + {\textstyle{5 \over 12}} n_F \right)
	{\alpha_s(\mu) \over \pi}
\;+\; \Bigg[ 244.9 - 17.24 n_F - 0.415 n_F^2
\nonumber \\
\;-\; {165 \over 8} \left( 1 + {\textstyle{5 \over 12}} n_F \right)
	\left( 1 - {\textstyle{2 \over 33}} n_F \right)
	\log {\mu \over 2 \pi T}
\;-\; 135 \left(1 + {\textstyle{1 \over 6}} n_F \right)
		\log {\Lambda_E \over 2 \pi T} \Bigg]
	\left( {\alpha_s \over \pi} \right)^2
\Bigg\} \;.
\label{fE}
\end{eqnarray}
This expression differs from that given in Ref.~\cite{solution},
where the counterterm (\ref{deltafE}) was not taken into account.

We have calculated two terms in the perturbation series for $m_E^2$
and three terms in the series for $f_E$.  We can use these results to
study the convergence of perturbation theory for the parameters
of EQCD.  We consider the case of $n_F=3$ flavors of quarks, although our
conclusions will not depend sensitively on $n_F$.  The question of
the convergence is complicated by the presence of the
arbitrary renormalization and factorization scales $\mu$ and $\Lambda_E$.
The next-to-leading-order (NLO) correction to $f_E$ is independent
of $\mu$ and $\Lambda_E$, and is small compared to the leading-order (LO)
term provided that $\alpha_s(\mu) \ll 1.1$.
The NLO correction to $m_E^2$ and the next-to-next-to-leading-order (NNLO)
correction to $f_E$ both depend on the renormalization scale $\mu$.
One scale-setting scheme that is physically well-motivated is
the BLM prescription \cite{BLM}, in which $\mu$ is adjusted to cancel
the highest power of $n_F$ in the correction term.
This prescription gives $\mu = 0.93 \, \pi T$ when applied to $m_E^2$
and $\mu = 4.4 \, \pi T$ when applied to $f_E$.  These values
differ only by about a factor of 2 from $2 \pi T$, which is the lowest
Matsubara frequency for gluons.  Below, we will consider the three values
$\mu = \pi T$, $2 \pi T$, and $4 \pi T$.
For the NLO correction to $m_E^2$ to be much smaller than the LO term,
we must have $\alpha_s(\mu) \ll 0.8$, 3.8, and 1.4 if
$\mu = \pi T$, $2 \pi T$, and $4 \pi T$, respectively.
Based on these results,
we conclude that the perturbation series for the parameters of
EQCD are well-behaved provided that $\alpha_s(2 \pi T) \ll 1$.

The NNLO correction for $f_E$ depends not only on $\mu$, but also on the
factorization scale $\Lambda_E$.  Because the coefficient of
$\log(\Lambda_E/2 \pi T)$
in (\ref{fE}) is so much larger than that of $\log(\mu/2 \pi T)$,
the NNLO correction for $f_E$
is much more sensitive to $\Lambda_E$ than to $\mu$.
It is useful intuitively to think of the
infrared cutoff $\Lambda_E$ as being much smaller than the
ultraviolet cutoff $\mu$.  However, these scales can be identified
with momentum cutoffs only up to multiplicative constants that may be
different for $\mu$ and $\Lambda_E$.  Both parameters are introduced
through dimensional regularization, but $\mu$ arises from ultraviolet
divergences of 4-dimensional integrals, while $\Lambda_E$ arises
from infrared divergences of 3-dimensional integrals.
We might be tempted to set $\Lambda_E = \mu$, but then the
NNLO coefficient in $f_E$ is large.  For the choice
$\mu = 2 \pi T$, the correction to the LO term is a multiplicative factor
$1 - 0.9 \alpha_s + 6.46 \alpha_s^2$.  The NNLO correction can be made small
by adjusting $\Lambda_E$.
It vanishes for $\Lambda_E = 5.8 \, \pi T$, $5.1 \,\pi T$,
and $4.5 \, \pi T$ if $\mu = \pi T$, $2 \pi T$, and $4 \pi T$, respectively.
We conclude that the perturbation series for $f_E$ is well-behaved
if the factorization scale $\Lambda_E$ is chosen to be approximately
$5 \pi T$.  Whether this choice is reasonable
can only be determined by calculating other EQCD parameters to higher order
to see if the same choice leads to well-behaved perturbation series.

The choice of $\Lambda_E$ that makes the perturbation series for the
EQCD parameters well-behaved may be much larger than the largest mass
scale $m_E$ of EQCD.  Perturbative corrections in EQCD will then
include large logarithms of $\Lambda_E/m_E$.  This problem can be avoided
by using renormalization group equations to evolve the parameters
of EQCD from the initial scale $\Lambda$ down to some scale $\Lambda_E'$
of order $m_E$.
The coefficient $f_E$ satisfies the renormalization group equation
\begin{equation}
\Lambda_E {d \ \over d \Lambda_E} f_E
\;=\; - {3 \over 2 \pi^2} g^2_E m^2_E \;.
\label{rgfE}
\end{equation}
The evolution of $g_E^2$ and $m_E^2$ occurs only at
higher order in the coupling constant and therefore can be ignored.
The solution to the renormalization group equation is therefore trivial:
\begin{equation}
f_E(\Lambda_E') \;=\; f_E(\Lambda_E)
- {3 \over 2 \pi^2} g^2_E m^2_E \log {\Lambda_E' \over \Lambda_E} \;.
\label{rgf}
\end{equation}

Having determined the parameters of EQCD to the necessary accuracy,
we proceed to calculate the free energy using (\ref{FEQCD}).
The contribution from the scale $T$ is given by the coefficient
$f_E$ in (\ref{fE}).  The contribution from the scale
$gT$ is given by $f_M = - \log {\cal Z}_{\rm EQCD}/T$.
In order to calculate $f_M$ using perturbation theory in EQCD,
we must include the effects of the mass parameter $m_E^2$ to all orders,
but the gauge coupling constant $g_E$ can be treated as a
perturbation parameter.
The contributions to $\log {\cal Z}_{\rm EQCD}$ of orders
$g^3$, $g^4$, and $g^5$ are given by the 1-loop, 2-loop, and 3-loop vacuum
diagrams in EQCD, respectively.
The integrals can be calculated analytically using
methods developed by Broadhurst \cite{broadhurst}.
The two-loop integrals include an ultraviolet pole in $\epsilon$
that is proportional to $g_E^2 m_E^2$.  This divergence is cancelled by
the counterterm $\delta f_E$ for the coefficient of the unit operator,
which is given in (\ref{deltafE}).
Our final result for the coefficient $f_M$ in (\ref{free}) is
\begin{equation}
f_M(\Lambda_E) \;=\; - {2 \over 3 \pi} m_E^3
\left[ 1 \;-\; \left( 0.256 - {9 \over 4} \log{\Lambda_E \over m_E} \right)
	{g_E^2 \over 2 \pi m_E}
	\;-\; 27.6 \left( {g_E^2 \over 2 \pi m_E} \right)^2
	\right] \;.
\label{fM}
\end{equation}
Note that the dependence of $f_M$ on $\Lambda_E$ cancels that
of $f_E$ in (\ref{fE}).  The expression (\ref{fM}) can be
expanded in powers of $g$ by setting $g_E^2 = g^2T$
and using the expansion (\ref{mE}) for $m_E^2$.

We now consider the convergence of the perturbation series (\ref{fM})
for $f_M$.  The size of the NLO correction depends on the choice of the
factorization scale $\Lambda_E$.  It is small if $\Lambda_E$ is chosen to be
approximately $m_E$.  The NNLO correction in (\ref{fM}) is independent
of any arbitrary scales.  If $n_F=3$, it is small compared to the leading order
term only if $\alpha_s \ll 0.17$.  Thus the perturbation series
for $f_M$ is well-behaved only for values of
$\alpha_s(2 \pi T)$ that are much smaller than those required
for the parameters of EQCD to have well-behaved perturbation series.

Adding (\ref{fE}) and (\ref{fM}) and expanding in powers of $\sqrt{\alpha_s}$,
the complete expression for the free energy $F$ is
\begin{equation}
F \;=\; - {8 \pi^2 \over 45} T^4
\left[ F_0
\;+\; F_2  {\alpha_s(\mu) \over \pi}
\;+\; F_3  \left( {\alpha_s(\mu) \over \pi} \right)^{3/2}
\;+\; F_4  \left( {\alpha_s \over \pi} \right)^2
\;+\; F_5  \left( {\alpha_s \over \pi} \right)^{5/2} \right] \;,
\label{freeg}
\end{equation}
where the truncation error is of order $\alpha_s^3 \log \alpha_s$.
The coefficients in this expansion are
\begin{eqnarray}
F_0 &=& 1 + \textstyle{21 \over 32} n_F \;,
\\
F_2 &=& - {15 \over 4} \left( 1 + \textstyle{5 \over 12} n_F \right) \;,
\\
F_3 &=& 30 \left(1 + \textstyle{1 \over 6} n_F \right)^{3/2} \;,
\\
F_4 &=&  237.2 + 15.97 n_F - 0.413 n_F^2
+ { 135 \over 2} \left( 1 + \textstyle{1 \over 6} n_F \right)
	\log \left( {\alpha_s \over \pi}
		\left(1 + \textstyle{1 \over 6} n_F \right) \right)
\nonumber \\
&& \;-\; { 165 \over 8}
\left( 1 + \textstyle{5 \over 12} n_F \right)
\left( 1 - \textstyle{2 \over 33} n_F \right)
	\log {\mu \over 2 \pi T} \;,
\\
F_5 &=&
\left( 1 + \textstyle{1 \over 6} n_F \right)^{1/2}
\Bigg[ -799.2 - 21.96 n_F - 1.926 n_F^2
\nonumber \\
&& \;+\; {495 \over 2} \left( 1 + \textstyle{1 \over 6} n_F \right)
	\left( 1 - \textstyle{2 \over 33} n_F \right)
		\log {\mu \over 2 \pi T} \Bigg] \;.
\label{F5}
\end{eqnarray}
The coefficient $F_2$ was first given by Shuryak \cite{shuryak},
while $F_3$ was first calculated correctly by Kapusta \cite{kapusta}.
The coefficient $F_4$ was computed in 1994
by Arnold and Zhai \cite{arnold}.
The coefficient $F_5$ in (\ref{F5}) has been calculated independently by
Kastening and Zhai using a different method \cite{zhai}.

We now ask how small $\alpha_s$ must be in order for the expansion
(\ref{freeg}) to be well-behaved.
For simplicity, we consider the case $n_F=3$,
although our conclusions are not sensitive to $n_F$.
If we choose the renormalization scale $\mu = 2 \pi T$ motivated
by the BLM criterion \cite{BLM},
the correction to the LO result is a multiplicative factor
$1 - 0.9 \alpha_s + 3.3 \alpha_s^{3/2}
	+ (7.1 + 3.5 \log \alpha_s) \alpha_s^2 - 20.8 \alpha_s^{5/2}$.
The $\alpha_s^{5/2}$ term is the largest correction unless
$\alpha_s(2 \pi T) < 0.12$.
We can make the $\alpha_s^{5/2}$ term small only by choosing the
renormalization scale to be near the value $\mu = 36.5 \pi T$
for which $F_5$ vanishes.  This ridiculously large of $\mu$ arises because
the scale $\mu$ has been adjusted to cancel the large $g^5$ correction
to $f_M$ in (\ref{fM}).  This contribution arises from the momentum
scale $gT$ and has nothing to do with renormalization of $\alpha_s$.
We conclude that the expansion (\ref{freeg}) for $F$ in powers of
$\sqrt{\alpha_s}$ is well-behaved only if $\alpha_s(2 \pi T)$
is an order of magnitude smaller than the value required for the
EQCD parameters to be well-behaved.

We now consider briefly the implications for theoretical studies
of the quark-gluon plasma.  We have found that the convergence of
perturbation theory requires much smaller values of $\alpha_s(2 \pi T)$
for quantities at the scale $gT$ than for quantities at the scale $T$.
The critical temperature $T_c$ for formation
of a quark-gluon plasma is approximately 200 MeV.  It may be possible
in heavy-ion collisions to produce a quark-gluon plasma with
temperatures several times $T_c$.  At $T = 350 \; {\rm MeV}$,
$\alpha_s(2 \pi T) \approx 0.3$, which is
small enough that perturbation theory
may be reasonably convergent at the scale $T$, but it is certainly
not convergent at the scale $gT$.  We conclude that at the temperatures
achievable in heavy-ion collisions, perturbative QCD may be accurate
when applied to quantities that involve the scale $T$ only.
However nonperturbative methods are required to accurately calculate
quantities that involve the scales
$gT$ and $g^2T$ associated with screening in the plasma.

This work was supported in part by the U.~S. Department of Energy,
Division of High Energy Physics, under Grant DE-FG02-91-ER40684,
and by the Ministerio de Educaci\'on y Ciencia of Spain.


\end{document}